\documentclass[a4paper,12pt, traditabstract, longauth]{article} 
\usepackage{jcappub} 
\usepackage{amsmath,amssymb,amsfonts,subfigure,epsfig}
\usepackage{epsfig}
\usepackage{dcolumn}
\usepackage{bm}
\usepackage{multirow}

\newcommand{\beq}{\begin{equation}}
\newcommand{\eeq}{\end{equation}}
\newcommand{\bea}{\begin{eqnarray}}
\newcommand{\eea}{\end{eqnarray}}

\newcommand{\phicmb}{\phi_{\text{CMB}}}
\begin{document}
\title{ Bound on largest $r\lesssim 0.1$ from sub-Planckian excursions of inflaton}

\author{Arindam Chatterjee$^{1}$}
\author{Anupam Mazumdar$^{2}$}

\affiliation{$^{1}$~Harish-Chandra Research Institute, Chhatnag Road, Jhusi, Allahabad, 211 019, India\\
$^{2}$~Physics Department, Lancaster University, LA1 4YB, UK}

\begin{abstract}
{In this paper we will discuss the range of large tensor to scalar ratio, $r$, obtainable from a sub-Planckian excursion of a {\it single}, 
{\it slow roll} driven inflaton field. In order to obtain a large $r$ for such a scenario one has to 
depart from a monotonic evolution of the slow roll parameters in such a way that one still satisfies all the current constraints
of \texttt{Planck}, such as the scalar amplitude, the tilt in the scalar power spectrum, running and running of the tilt
close to the pivot scale. Since the slow roll parameters evolve non-monotonically, we will also consider the evolution of the power spectrum
on the smallest scales, i.e. at ${\cal P}_{s}(k\sim 10^{16}~{\rm Mpc^{-1}})\lesssim 10^{-2}$, to make sure that the amplitude does not become too 
large. All these constraints tend to keep the tensor to scalar ratio, $r\lesssim 0.1$. We scan three different kinds of potential for supersymmetric flat directions 
and obtain the benchmark points which satisfy all the constraints.
We also show that it is possible to go beyond $r\gtrsim 0.1$ provided we relax the upper bound on the power spectrum on the smallest scales.}
\end{abstract}

\maketitle

\section{Introduction}

The recent observation of B-mode polarization in the Cosmic Microwave Background 
Radiation (CMBR) at large scales by \texttt{BICEP} has generated a lot of interest \cite{Bicep2}. 
Assuming the origin of these B-modes is primordial tensor perturbation generated during inflation~\footnote{It is possible to select a 
gauge where the tensor perturbations can be solely generated from pure classical gravity~\cite{Ashoorioon:2012kh}. However, the 
classical gravity cannot yield a large tensor-to-scalar ratio with classical initial conditions. One must assume quantum initial conditions for the 
tensor modes to be observable on the largest angular scales, see~\cite{Ashoorioon:2012kh}. } the tensor-to-scalar 
ratio $r=0.20^{+0.07}_{-0.05}$ \cite{Bicep2}, which is in tension with the bound from \texttt{Planck} \cite{Planck}. 
 However, this claim has been debated \cite{MMUS}, and it has been pointed out that dust might contribute 
significantly in the B-mode polarization in the region of interest for \texttt{BICEP}. 
The complete dust map, once released by \texttt{Planck} collaboration, will shed light 
on this issue. 

Generically, a large value of $r$ poses several problems in the context of single 
field slow-roll inflation models, for a review on inflation, see~\cite{Mazumdar:2010sa}. In particular, models in which the inflaton 
undergoes a sub-Planckian excursion during inflation, it is difficult to realize 
a large $r$. The well-known Lyth bound gives, \cite{Lyth}, 
\begin{equation}\label{Lyth}
\frac{{\cal P}_{t}}{{\cal P}_{s}}\equiv  r \lesssim 3 \times 10^{-3} \left(\frac{50}{N}\right)^2 \left(\frac{\Delta \phi}{M_P}\right)\,,
\end{equation}
where, ${\cal P}_{s}$  and ${\cal P}_{t}$ are scalar and tensor power spectra.
Assuming a field excursion $\Delta \phi \lesssim M_P\sim 2.4\times 10^{18}$~GeV~\footnote{For the validity of an effective field
theory, we would require that at the pivot scale, $\phi_{CMB}\lesssim M_p$, and $\Delta\phi < M_P$.} during inflation and if it 
generates $N\sim 50$ e-foldings of inflation, this constraints $r \lesssim 0.003$ from Eq.~(\ref{Lyth})~\footnote{This simple bound
gets modified if there were large number of fields participating during inflation, as in the case of {\it Assisted inflation}~\cite{Liddle:1998jc,Kanti:1999ie,Jokinen:2004bp}. One would be 
able to obtain large $r$ with sub-Planckian field displacements, but at the cost of large number of fields, i.e. ${\cal O}(10^{3} -10^6)$.}.

However, in  \cite{HGRS,Ben,HMN} it has been argued that the Lyth bound can be evaded if 
non-monotonic evolution of the slow-roll parameters is assumed and/or if 
the field generating the perturbations does not require to produce the entire 
efoldings of expansion \cite{HMN,Lowinf,Antusch:2014cpa}. In particular, if it is required to produce sufficient 
expansion spanning about $8$ efoldings of the observational window (around the relevant 
CMB pivot scale), the CMB constraints can be satisfied, and it is possible to obtain 
$r \lesssim 0.1$. 

Although the inference from the \texttt{BICEP} data has been debated, it remains 
interesting to investigate how large $r$ can be accommodated in the small field 
inflation models. In the 
present work, we consider simple renormalizable and non-renormalizable potentials
to show that it is possible to obtain large value of $r\lesssim 0.1$ if we satisfy all 
the current constraints arising from \texttt{Planck} and assuming that on 
small scales the scalar power spectrum does not become too large. For the purpose
of our discussion, we will assume that on the smallest scales, ie. ${\cal P}_{s}(k\sim 10^{16}~{\rm Mpc^{-1}})\lesssim 10^{-2}$.
There is no hard and fast bound  on small scales, this bound primarily arises from the creation of 
primordial blackholes (PBHs), which are formed due to the collapse of matter on small scales when the matter perturbations 
reach the contrast $\delta\rho/\rho \sim {\cal O}(0.1)$, or so~\cite{CKSY,DE1}. In principle, there is a scope for improving this bound further
and therefore relaxing some of our own constraints on the value of $r$. It is also probable that we do not form PBHs due to short distance modifications
of gravity~\cite{Biswas:2011ar}, in which case one might be able to relax some of the constraints at low scales, we will speculate on this possibility too.

We will work in a simple framework of sub-Planckian inflation, first with all possible allowed renormalizable contributions 
in the potential, such as $V_0,~\phi^2,~\phi^3, ~\phi^4$, which can be motivated with or without supersymmetry (SUSY).
Then we would investigate some non-reanormalizable contributions to the potential, and we would motivate them from 
supersymmetric Standard Model (SUSY-SM) {\it gauge invariant} $D$-flat directions~\cite{Enqvist:2003gh}. The flat directions are 
natural consequences of SUSY, in particular we will discuss the potentials
arising from SUSY partners of quarks and leptons, and the SUSY Higgses themselves~\footnote{There are many issues with the 
validity of an effective field theory for super-Planckian excursions. The potential term obtains
non-renormalizable corrections as well as the kinetic terms obtain higher derivative corrections. It may be possible to suppress the
non-renormalizable potential corrections via invoking some symmetry such as shift symmetry~\cite{Kawasaki:2000yn,Mazumdar:2014bna}, but higher derivative corrections
which obey shift symmetry cannot be neglected, and pose potential problems such as {\it ghosts, instabilities}, etc.~\cite{Chialva:2014rla}, and typically one requires a full
UV understanding of gravity and the inflaton sector, which we currently lack at the moment. String theory is not a UV complete theory, yet.
It is believed that embedding inflation within string theory will ameliorate some of these issues, but there are many challenges such as $\alpha'$ 
corrections in the gravity sector and in the matter sector, which are poorly understood within string theory, see ~\cite{Chialva:2014rla}.}. In all of these examples, the presence of 
large cosmological constant, $V_0$, will play a central role. We will assume
$V^{1/4}_0\sim 10^{16}$~GeV, in order to produce large detectable $r\sim 0.1$.

The article is organised as follows. In section \ref{sec:pot} we describe the potential, 
and the general constraints on the potential imposed in the context of inflation. We use 
the measured values of the amplitude of the scalar perturbation and the spectral index 
to reconstruct the potential at the pivot scale ($k_{pivot}$) and then present relevant 
benchmark points with large $r$ (at $k_{pivot}$). In section \ref{sec:pert} we study the 
scale dependence of the tensor and scalar perturbations for our benchmark scenarios. 
Finally we conclude in \ref{sec:concl}.


\section{The Potential, Constraints and Large $r$}
\label{sec:pot}

\subsection{Motivating the potentials}
The most general renormalizable potential should contain terms:
\begin{eqnarray}
V(\phi)=  V_0+ A \phi^2 - B \phi^3 + C \phi^{4}.
\label{eq:pot-0}
\end{eqnarray}
where $V_0$ is the cosmological constant, $A$ is the quadratic mass term and $B$ has the mass dimension and $C$ is 
dimensionless parameter. As such the $\phi$ field is unidentified and could be treated here as a singlet. Note that the sign in front of $B$ is negative,
because if all $A, B, C$ were positive we would not be able to construct a {\it flat potential} below the VEV $M_P$. In order to construct a flat potential
below $M_P$, we would need either a saddle or inflection point, which can be obtained if we make a simple choice of  sign infront of $B$ is negative~\cite{Allahverdi:2006cx}.
Further note that the above potential can also be embedded within a physical setup within SUSY and in particular when we consider supergravity (SUGRA) corrections.

Within SUSY, for a generic superpotential:
\begin{equation}
W=\lambda\frac{\Phi^{n}}{M_P^{n-3}}
\end{equation}\label{supot}
where $\Phi$ is flat direction superfield, and $\lambda$ is the self coupling whose value simply shifts the VEV of the flat direction, 
and in this paper the sliding VEV of  $\phi$ is always below $M_{P}$. We will be considering $n\gtrsim 3$, for $n=3$, the renormalizable
potential will take the form of Eq.~(\ref{eq:pot-0}). The $\phi$ field obtains a soft SUSY breaking mass term 
$m_{\phi}\gtrsim {\cal O}(1~{\rm TeV})$. Together with the non-renormalizable operator, this gives a potential for $\phi$ which is 
determined {\it solely} by $n$ and $m_{\phi}$.

Within  $N=1$ SUGRA, the  potential is typically dominated by the Hubble induced corrections, namely
the Hubble induced mass correction and the Hubble induced $A$ term, see~\cite{Dine:1995uk}.  Assuming for the time being {\it canonical K\"ahler potential}
for the inflaton field and any other heavy fields are static during the inflationary phase, the potential can be recast in 
with $M_P^2 = 1$, in the following form (for a detailed derivation from $N=1$ SUGRA potential, see~\cite{Mazumdar:2011ih,Choudhury:2014sxa,KK}, 
for a review see~\cite{Mazumdar:2010sa,Masahide}): 
\begin{eqnarray}
V(\phi)&=& V_0+ c_H H^2 \phi^2 - a_H H \lambda_n \phi^n + \lambda_n^2 \phi^{2n-2}\,, \label{eq:pot1}\\
       &=& V_0+ A \phi^2 - B \phi^n + C \phi^{2n-2}.
\label{eq:pot}
\end{eqnarray}
where $V_0\approx3H^{2}M_{P}^{2}$ and $H$ is the Hubble parameter during inflation. There could be many sources of $V_0$, a string theory landscape~\cite{Douglas:2006es}, 
or within SSM landscape~\cite{MN-curvaton,Allahverdi:2008bt}, or there could be hidden sector contributions~\cite{Enqvist:2007tf,Lalak:2007rsa}, or there could be a combination of these effects.  Similarly, a hybrid model of inflation~\cite{Linde:1993cn} also provides a  constant vacuum energy density during inflation. For an example, we may consider  the hidden 
sector superpotential, $W=M^{2}I,$ where $M$ is some high scale which dictates the initial vacuum energy density, $V_0$, and $I$ is the heavy superfield. One may also consider:
$W=\phi(I^{2}-M^{2})$, for details, see~\cite{Mazumdar:2011ih,Choudhury:2014sxa}. In our case, we will assume that the the heavy field $I$ is dynamically settled in its vacuum and static when the relevant perturbations leave the Hubble patch during inflation.

The second term is the Hubble-induced mass term~\cite{Bertolami:1987xb,Dine:1995uk}. The coefficient $c_{H}$ depends on the exact nature of the K\"ahler potential and it can also absorb the higher order K\"ahler corrections~\cite{Dine:1995uk}.  Typically, this term can ruin the flatness, as $m_{\phi}^{2}\ll c_{H}H^{2}$. For large $c_{H}\sim{\cal O}(1)$, the potential is $V\approx 3H^{2}M_{P}^{2}+H^{2}\phi^{2}+...$, which leads to the slow-roll parameter $\eta=M_{P}^{2}V''/V=c_{H}\sim {\cal O}(1)$. The third term is the Hubble-induced $A$-term,  where the coefficient $a_{H}$ is a positive dimensionless and of order $\sim c_H$.  However, combining all the terms in Eq.~(\ref{eq:pot1}), one can drive inflation as shown in a number of papers due to the presence of saddle or inflection point~\cite{Mazumdar:2011ih,Choudhury:2014sxa}. Note that we have categorically selected the sign in front of $c_H$ positive while in front of $a_H$ negative. We could have chosen vice-versa without any loss of generality, but we will stick to our current notation. While selection $c_H<0$, we can also address the issue of large cosmological constant $V_0$ which may arise 

Let us now provide very briefly inflaton candidates for $n=3,~4$, and $6$.

\begin{enumerate}
\item{$n=3$ case:  The simplest candidate belongs to the superpotential term: $ W\sim \lambda NH_uL$ flat direction of supersymmetric Standard Model  (SSM)$\times U(1)_{B-L}$, 
where $U(1)_{B-L}$ is gauged. This is a $D$-flat direction where $N$ is the right handed neutrino component, $H_u$ is one of the Higgses which give masses to up type quarks
and leptons, and $L$ corresponds to left handed lepton. The inflaton field is given by~\cite{Allahverdi:2006cx}:
\begin{equation}
\phi=\frac{\widetilde N+ H_{u} +\widetilde L}{\sqrt{3}}\,,
\end{equation}
where $\widetilde N$ is the right handed neutrino component and $\widetilde L$ corresponds to the left handed slepton.
 }

\item{ $n=4$ case: The simplest example for this case is the SSM Higgses. It is known that the $H_uH_d$ can support inflection point inflation~\cite{MSSMH}, 
with the superpotential contribution: $W\sim (H_u H_d)^2/M_P$, where $H_d$ provide masses to down type quarks and leptons, and $\mu\sim {\cal O}(1)$~TeV. 
The inflaton field is given by:
\begin{equation}
\phi=\frac{ H_{u} + H_d}{\sqrt{2}}\,.
\end{equation}
}

\item{ $n=6$ case:  This case has been studied very extensively in Refs.~\cite{Allahverdi:2006iq}. The two well known inflaton candidates are $W\sim \lambda (LLe)^2/M_P^3$
comprising the leptonic superfields with $e$ corresponds to right handed electron superfield, and the one with $\lambda (udd)^3/M_P^3$, comprising the right handed
quarks. The inflaton fields are given by:
\begin{equation}
\phi=\frac{ \widetilde u+ \widetilde d +\widetilde d}{\sqrt{3}}\,,~~~~~~~~~~~\phi=\frac{ \widetilde L+ \widetilde L +\widetilde e}{\sqrt{3}}\,,
\end{equation}
where $\widetilde u,~\widetilde d$ correspond to right handed squarks and $\widetilde L$ corresponds to left handed slepton, and $\widetilde e$ corresponds to selection.
}
\end{enumerate}

We will be analysing all the three possibilities in the rest of the paper and will see which of the directions can yield the largest possible $r$.

\subsection{Slow roll parameters and observables}

The potential must satisfy slow-roll conditions to give rise to the  exponential expansion 
during inflation. This requires the smallness of the following slow-roll parameters: 
\begin{eqnarray}\label{eq:slowroll0}
\varepsilon_V= \frac{M_P^2}{2}\left(\frac{V'}{V}\right)^2; 
~\eta_V=M_P^2 \left(\frac{V''}{V}\right); \\
\xi^2_V=M_P^4 \left(\frac{V'V'''}{V^2}\right);
~\sigma^3_V=M_P^6 \left(\frac{V'^2 V''''}{V^3}\right). 
\label{eq:slowroll}
\end{eqnarray}
The parameters determined by \texttt{CMB} observations are related to these slow-roll 
parameters as follows, 
\begin{align}
A_\mathrm{s} & \approx  \frac{V}{24 \pi^2 M_\mathrm{pl}^4 \varepsilon_V}, \label{eq:as_def} \\
n_\mathrm{s} -1 &\approx 2 \eta _V- 6 \varepsilon _V, \label{eq:ns_def} \\
\mathrm{d}n_\mathrm{s}/\mathrm{d}\ln k & \approx   16 \varepsilon _V\eta _V- 24 \varepsilon^2_V - 2 \xi^2_V, 
\label{eq:alphas_def} \\
\mathrm{d}^2n_\mathrm{s}/\mathrm{d}\ln k^2  & \approx  -192 \varepsilon^3_V + 192 \varepsilon^2_V \eta _V- 
32 \varepsilon _V\eta^2_V \\  
& \quad - 24 \varepsilon _V\xi^2_V + 2 \eta _V\xi^2_V + 2 \sigma^3_V \label{eq:betas_def},\\
A_\mathrm{t} & \approx  \frac{2 V}{3 \pi^2 M_\mathrm{pl}^4}, \label{eq:at_def} \\
n_\mathrm{t} &\approx - 2 \varepsilon_V,\\
\mathrm{d}n_\mathrm{t}/\mathrm{d}\ln k& \approx    4 \varepsilon_V \eta_V - 8 \varepsilon^2_V. \label{eq:alphat_def} 
\end{align}
In the above set of equations, $A_s$ and $n_s$ denote the amplitude of the scalar perturbation and 
the scalar spectral-tilt respectively; while $A_t$ and $n_t$ denote the same for tensor perturbations. 

Assuming that the subsequent slow-roll parameters, and therefore the scale dependence of 
$\mathrm{d}^2n_\mathrm{s}/\mathrm{d}\ln k^2$ are much smaller, the power-spectrum for the tensor 
and the scalar perturbations can be expressed as, \footnote{At Hubble crossing $k= a H$.} 
\begin{align}
\mathcal{P}_{s}(k) & = \dfrac{1}{8 \pi^2} \dfrac{ H^2}{\varepsilon_V} \Big|_{k= aH} \\ &=  A_\mathrm{s} \left( \frac{k}{k_*}\right)^{n_\mathrm{s}-1 + 
\frac{1}{2} \, \mathrm{d}n_\mathrm{s}/\mathrm{d}\ln k \ln(k/k_*) + \frac{1}{6} \, 
\mathrm{d}^2n_\mathrm{s}/\mathrm{d}\ln k^2  \left( \ln(k/k_*) \right)^2 + ...}, \label{scalarps}\\
\mathcal{P}_\mathrm{t}(k)  &= \dfrac{2 H^2}{\pi^2} \Big|_{k= aH} \\ &= A_\mathrm{t} \left( \frac{k}{k_*}\right)^{n_\mathrm{t} + \frac{1}{2} \, 
\mathrm{d}n_\mathrm{t}/\mathrm{d}\ln k \ln(k/k_*) + ... }, 
\label{eq:ptps}
\end{align}
where $a$ and $H$ denotes the scale factor and the Hubble parameter respectively,  
$k_*$ denotes a reference scale such that $\mathcal{P}_{s}(k_*)= A_s$. 
$k_*$ is assumed to be the \textit{pivot} ($k_{pivot}$) scale for comparison with the \texttt{Planck} data. 
Note that, in case of non-monotonicity of $\varepsilon$, however, such a simple power-law 
dependence may not hold good \cite{HMN}. As we will discuss in some detail, we will require to solve 
the full Mukhanov-Sasaki equation \cite{MFB} without slow-roll approximation for our benchmark 
values in the subsequent section~\footnote{We have compared that the analytical approximations work well
on the  largest scales, i.e. at the pivot scale and up to $k\sim 1{\rm Mpc^{-1}}$, but below this scale the analytical expressions
break down, therefore it is paramount that we evolve the perturbed mode equations all the way from the largest scales to the small
scale by using the Mukhanov-Sasaki  variable, for details see~\cite{MFB}. This point was first highlighted in Ref.~\cite{HMN} in the context 
of non-monotonic evolution of slow roll parameters.}.

The tensor-to-scalar ratio $r$ at the pivot scale is given by~\footnote{By definition we will always read the value of $r$ at the pivot scale, $k=k_\ast$.}, 
\beq
r(k=k_\ast) = \dfrac{\mathcal{P}_{t}(k_\ast)}{\mathcal{P}_{s}(k_\ast)}. 
\label{eq:r}
\eeq
This can be expressed accurately as \cite{LYTH1},  
\begin{align}
r & = 16 \varepsilon_{V}  \frac{\left[1-({\cal C}_{E}+1)\varepsilon_{V}\right]^{2}}
{\left[1-(3{\cal C}_{E}+1)\varepsilon_{V}+{\cal C}_{E}\eta_{V}\right]^{2}} \\ 
& \simeq 16 \varepsilon_V, 
\label{eqr}
\end{align}
where $\varepsilon_V,~\eta_V$ are the slow-roll parameters; ${\cal C}_{E}=4(\ln 2+\gamma_{E})-5$ 
with $\gamma_{E}=0.5772$ is the Euler-Mascheroni constant. 

Since the inflaton potential must satisfy all constraints from the CMB observations, 
we first discuss the updated constraints from \texttt{Planck} in the next subsection, 
before getting into the discussion on obtaining large $r$.

\begin{table*}[t]
\centering
\begin{tabular}{c|c|cc}
\hline
\hline
Model & Parameter &  {\it Planck}+WP+lensing & {\it Planck}+WP+high-$\ell$  \\
\hline
\hline
\multirow{7}{*}{\phantom{$\Bigg|$}+ $\mathrm{d}^2 n_\mathrm{s}/\mathrm{d} \ln k^2$} 
\phantom{$\Big|$} &\phantom{$\Big|$}
$n_{\mathrm{s}}$  & $0.9573^{+0.077}_{-0.079}$ & $0.9476^{+0.086}_{-0.088}$ \\
{$\Lambda$CDM + $\mathrm{d} n_\mathrm{s}/\mathrm{d} \ln k$}& \phantom{$\Big|$}
$\mathrm{d} n_\mathrm{s}/\mathrm{d} \ln k$ & $0.006^{+0.015}_{-0.014}$ 
& $0.001^{+0.013}_{-0.014}$ \phantom{$\Big|$}\\
& $\mathrm{d}^2 n_\mathrm{s}/\mathrm{d} \ln k^2$ \phantom{$\Bigg|$}
& $0.019^{+0.018}_{-0.014}$ & $0.022^{+0.016}_{-0.013}$\phantom{$\Big|$} \\
\hline
\multirow{6}{*}{$\Lambda$CDM + $r$ + $\mathrm{d} n_\mathrm{s}/\mathrm{d} \ln k$}
& $n_{\mathrm{s}}$ &  $0.9633 \pm  0.0072$ & $0.9570 \pm 0.0075$  \phantom{$\Big|$}\\
& $r$ &  $<0.26$ & $<0.23$\\
& $\mathrm{d} n_\mathrm{s}/\mathrm{d} \ln k$ & 
 $-0.017 \pm 0.012$ & $-0.022^{+0.011}_{- 0.010}$  \phantom{$\Big|$}\\
\hline
\end{tabular}
\vspace{.2cm}
\caption{\label{tab:cons}
Constraints on the primordial perturbation parameters for 
$\Lambda$CDM+
$\mathrm{d} n_\mathrm{s}/\mathrm{d} \ln k$+$r,$ and 
$\Lambda$CDM+$\mathrm{d} n_\mathrm{s}/\mathrm{d} \ln k$+
$\mathrm{d}^2 n_\mathrm{s}/\mathrm{d} \ln k^2$ models from \texttt{Planck} 
combined with other data sets. Constraints on the spectral index and its 
dependence on $k$ are given at the pivot scale of 
$k_* = 0.05$~Mpc$^{-1}$. The Table has been adopted from \cite{Planck}.} 
\end{table*}

\subsection{Observable constraints} 

Constraints from \texttt{Planck} have been summarised in Table \ref{tab:cons}
\cite{Planck}.

Note that there is no analyses considering 
\texttt{$\Lambda \text{CDM}+ \frac{d n_s}{d \ln k}+\frac{d^2 n_s}{d \ln k^2}+ r $}.  
\texttt{$\Lambda \text{CDM}+\frac{d n_s}{d \ln k}$} and either
$\frac{d^2 n_s}{d \ln k^2}$ or $r$  have been considered  for  the 
potentials given in Eqs.~(\ref{eq:pot1},\ref{eq:pot}). The constraints apply at the \textit{pivot} scale $k_*= 0.05~ \text{Mpc}^{-1}$. 
In our benchmark scenarios, therefore, we require that the parameters $\frac{d n_s}{d \ln k}$, 
$\frac{d^2 n_s}{d \ln k^2}$ and $r$ are within the range specified in Table \ref{tab:cons}.
Considering both $r$ and running, \texttt{Planck+WP} implications for slow-roll 
parameters are $\varepsilon_V < 0.015$ at 95\% CL, $\eta_V = -0.014_{-0.011}^{+0.015}$, 
and $|\xi^2_V| = 0.009 \pm 0.006$ \cite{Planck}. 
 \footnote{Considering only \texttt{$\Lambda\text{CDM}$} and $r$, a similar 
constraint on $n_s$ and $r_{0.002}$ has been depicted in Table 4 of \cite{Planck}.}
Considering only \texttt{Planck + WP} data, and assuming \texttt{$\Lambda \text{CDM}$} 
model, the following constraint on $A_s$ holds : 
$\ln (10^{10} A_{s*}) = 3.089^{+0.024}_{-0.027}$.
From \texttt{BICEP2} \cite{Bicep2}, we have, $ r(k_\ast)=  0.20^{+0.07}_{-0.05}$.  
This constraint may be relaxed once uncertainties in the estimation of 
the galactic foreground emission is considered \cite{MMUS}. Also, it has been shown 
that large power in small length scale may produce primordial black holes \cite{CKSY, DE1}. 
To avoid this, we impose ${\cal P}_s (k \simeq 10^{16}{\rm Mpc}^{-1}) \lesssim 10^{-2}$ for our benchmark scenarios. 


\subsection{Reconstruction of potential in terms of $r$}

In this subsection, we attempt to construct the potential, as expressed 
in Eq.~(\ref{eq:pot}), such that it is consistent with all the observational 
constraints. We follow the approach given in Ref.~\cite{HMN}. 

Let the \textit{pivot} scale $(k_*)$ correspond to  $\phicmb=1~M_P$. Then, for a given value of 
$\phicmb$ and $V_0$, we obtain a matrix equation for the variables $A$, $B$ and $C$: 

\begin{figure}[t!]
\epsfig{file=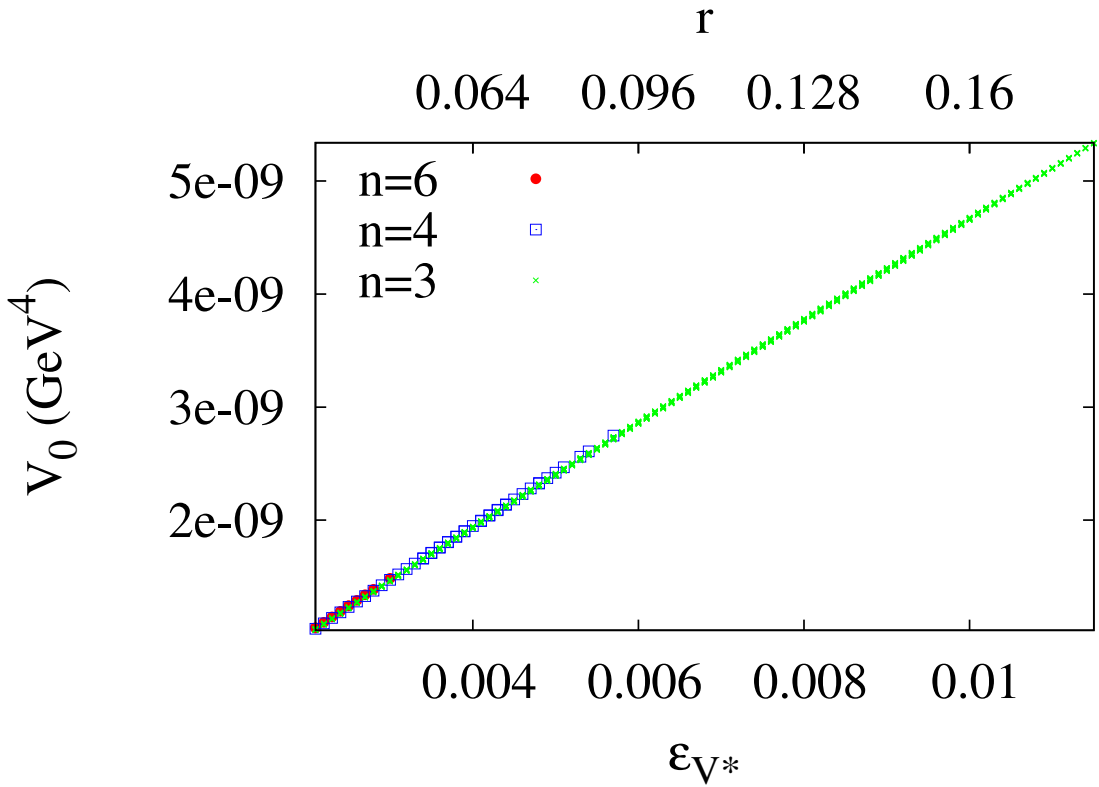, width= \textwidth}
\caption{The values of $V_0$ and $\varepsilon_{V*}$, for which  the potentials 
at $\phicmb = M_P$ satisfy all the CMB constraints at the \textit{pivot} scale 
have been shown for $n \in \{3,4,6\}$. }
\label{fig:0}
\end{figure}
\begin{figure}[t!]
\epsfig{file=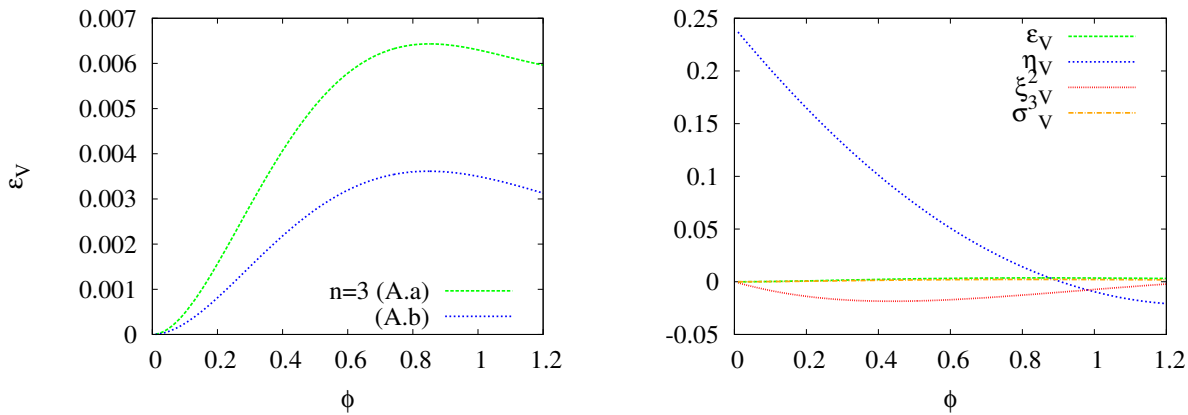, width= \textwidth}
\caption{In the left panel evolution of $\varepsilon_V$ with $\phi$ has been shown for benchmark 
scenarios (A.$a$) and (A.$b$) for $n=3$ case, see Table 2. In the right panel the slow-roll parameters, 
corresponding to the benchmark (A.$a$) have been plotted against $\phi$ using 
Eqs.~(\ref{eq:slowroll0},\ref{eq:slowroll}).}
\label{fig:1}
\end{figure}

\begin{figure}[ht!]
\epsfig{file=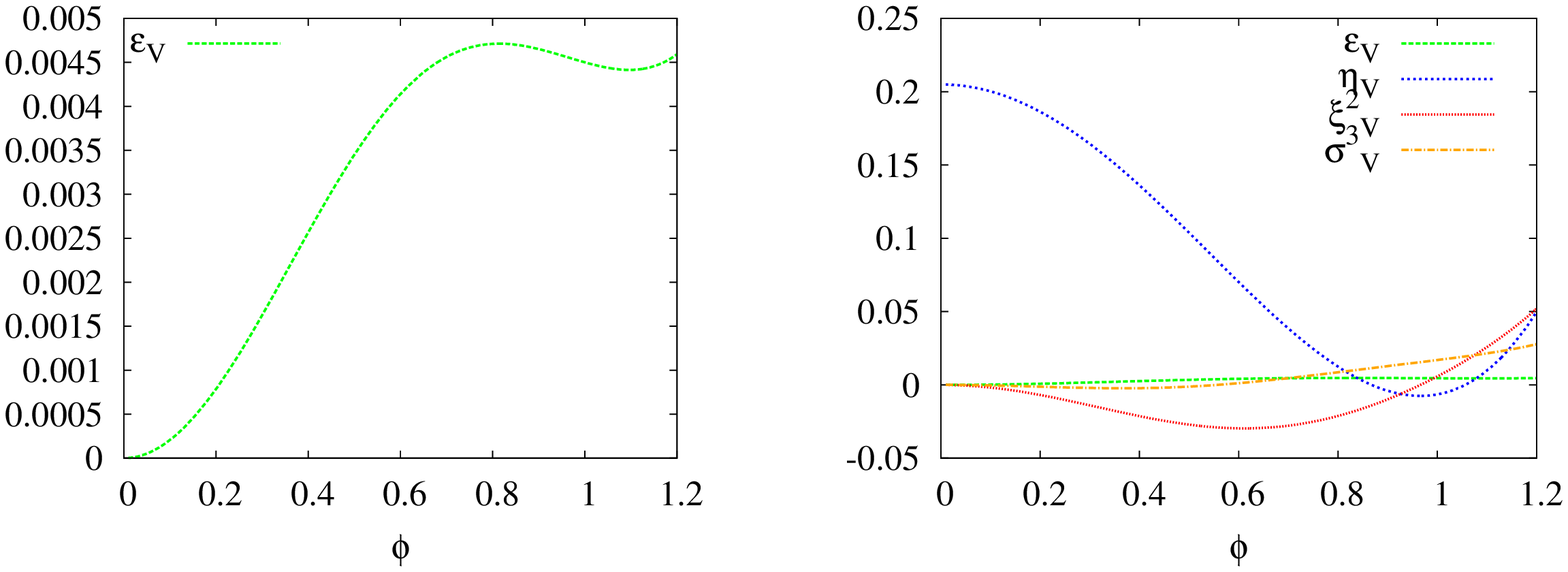, width= \textwidth}
\caption{The slow-roll parameters, corresponding to the benchmark for n=4 case, 
have been plotted against $\phi$ using Eqs.~(\ref{eq:slowroll0},\ref{eq:slowroll}).}
\label{fig:1a}
\end{figure}

\begin{figure}[t!]
\epsfig{file=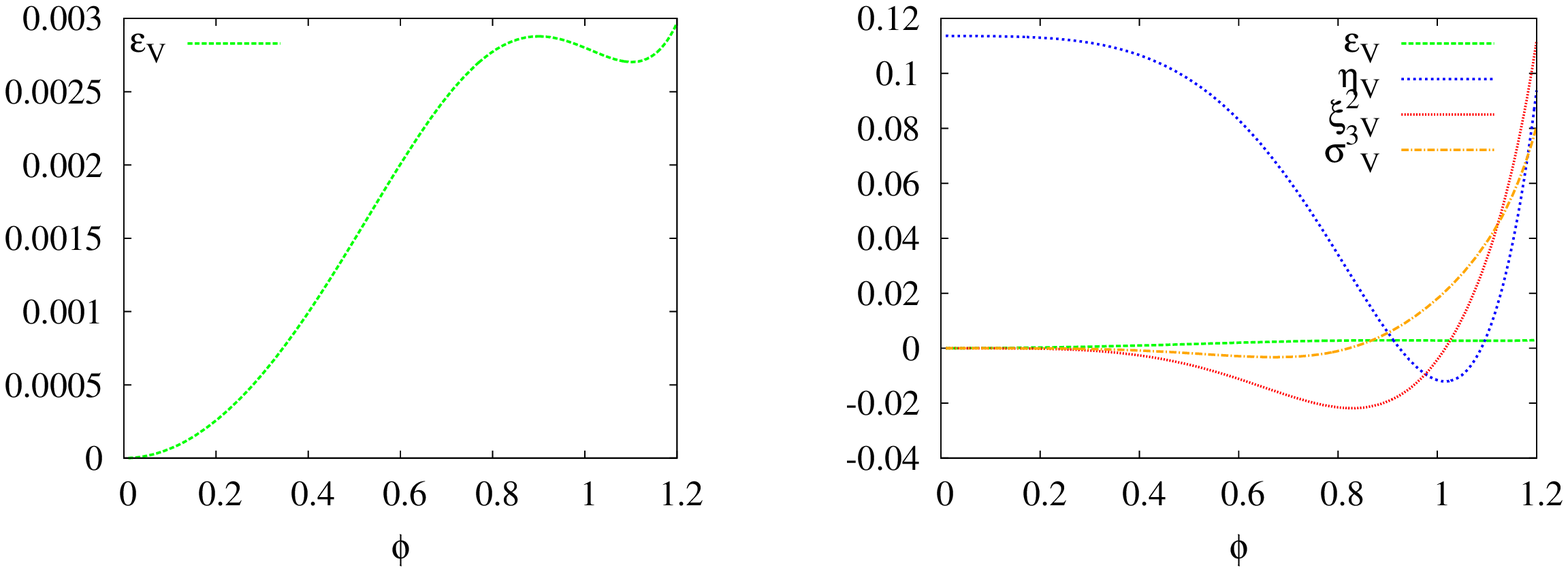, width= \textwidth}
\caption{The slow-roll parameters, corresponding to the benchmark for n=6 case, 
have been plotted against $\phi$ using Eqs. (\ref{eq:slowroll0},\ref{eq:slowroll}).}
\label{fig:1b}
\end{figure}

\begin{equation}
\left(\begin{tabular}{ccc}
$\phicmb^2$  & $-\phicmb^n$ & $\phicmb^{2(n-1)}$\\
$2\phicmb$ & $-n\phicmb^{(n-1)}$ & $2(n-1)\phicmb^{2n-3}$\\
$2$ & $-n(n-1)\phicmb^{(n-2)}$ & $2(n-1)(2n-3)\phicmb^{2(n-2)}$
      \end{tabular}\right)
\left(\begin{tabular}{c}
      $A$\\
      $B$\\
      $C$\\
      \end{tabular}\right)
\begin{tabular}{c}
      \\
      =\\
      \\
      \end{tabular}
\left(\begin{tabular}{c}
      $V(\phicmb)-V_0$\\
      $V'(\phicmb)$\\
      $V''(\phicmb)$\\
      \end{tabular}\right).
\label{eq:abc}
\end{equation}
Eqs. (\ref{eq:as_def}), (\ref{eq:ns_def}) and (\ref{eqr}) can be used to express 
$ V(\phicmb),~ V'(\phicmb),~ V''(\phicmb)$ in terms of observable parameters at 
the \textit{pivot} scale. This gives, 
\bea
V(\phicmb) &= &\frac{3}{2} A_s r \pi^2, \\
V'(\phicmb) &=& \frac{3}{2} \sqrt{\frac{r}{8}}(A_s r \pi^2), \\
V''(\phicmb) &=& \frac{3}{4} \left(\frac{3r}{8}+n_s-1\right)(A_s r \pi^2). 
\label{eq:recon}
\eea
Then Eqs. (\ref{eq:abc}) is used to determine the coefficients $A,~B$ and $C$ 
in terms of $V_0,~\phicmb$ and $r$. In the above equations, we have used 
$n_s = 0.96$ and $A_s = 2.2 \times 10^{-9}$.
Since we are interested in obtaining a large $r$ with sub-Planckian field excursion, 
we consider the maximum possible $\phicmb = M_P$ consistent with our requirement. 
We further require the potential to generate 50 efoldings of expansion, i.e. $N=50$. 
Expressing $r$ in terms of $\varepsilon_{V*}$ using Eq. (\ref{eq:r}), we scan over 
$\{V_0,~\varepsilon_{V*}\}$. The values of $\{V_0,~\varepsilon_{V*}\}$, for which the 
potential satisfies all CMB constraints at the \textit{pivot} scale, have been shown in 
figure (\ref{fig:0}). The red, blue and green dots correspond to $n=6$, $n=4$ and $n=4$ 
cases respectively. Note that, there is no assurance that for these values of $V_0$ 
and $\varepsilon_{V*}$ the power spectrum will satisfy all constraints at small length 
scales. We present our benchmark values in Table (\ref{tab:bm}). For three of our benchmark 
scenarios we assure that ${\cal P}_s (k \simeq 10^{16}{\rm Mpc}^{-1}) \lesssim 10^{-2}$. 
 
\begin{table}[ht!]
\begin{center}
\begin{tabular}{|c|c|c|c|c|c|c|  } \hline
     & $n$ & $V_0$  & $A$ & $B$ & $C$ & $r$ \\
\hline
\hline
(A.$a$) & 3 &  $3.010 \times 10^{-9}$ & $ 5.114 \times 10^{-10}$  & $3.126 \times 10^{-10}$& $7.075 \times 10^{-11}$& 0.101  \\ 
(A.$b$) & 3 &  $1.710\times 10^{-9}$ & $ 2.061 \times 10^{-10}$  & $1.165 \times 10^{-10}$& $2.248 \times 10^{-11}$& 0.056 \\ 
(B) & 4 & $2.184\times 10^{-9}$  & $2.238\times 10^{-10}$ & $ 8.301 \times 10^{-11}$  & $1.778 \times 10^{-11}$ & 0.072 \\ 
(C) & 6 & $1.388 \times 10^{-9}$ & $7.884\times 10^{-11}$ & $1.096 \times 10^{-11} $& $1.712 \times 10^{-12}$  & 0.045  \\    
\hline
\end{tabular}
\end{center}
\caption{Rows (A), (B) and (C) demonstrate benchmark points for scenarios (A), (B) and (C) respectively. The estimates
for $r$ have been rounded off to 3 decimal places. Read the values of $V_0,~A,~B,~C$  in terms of $M_P=1$ unit.}
\label{tab:bm}
\end{table}

\begin{itemize}
\item {\bf Benchmark A }: (n = 3 case) \\
The parameters of the potential has been specified in row 1 and 2 of Table \ref{tab:bm}. 

\subitem (a)
In (A.$a$), we relax the criteria that at short length scales $(k \simeq 10^{16} {\rm Mpc}^{-1})$ 
the scalar-power spectrum should be less than $10^{-2}$; we assume 
$P_s(k \sim 10^{14} {\rm Mpc}^{-1}) \sim \mathcal{O}(1)$ instead. We focus on obtaining $r$ 
as large as $0.1$, satisfying the constraints imposed at the \textit{pivot} scale. 
The slow-roll parameters at $\phicmb$ are given by, 
\beq
\varepsilon_{V*} = 0.0063, ~\eta_{V*}= -0.0011, 
~\xi_{V*}^2= -0.0061, ~\sigma_{V*}^3= 0.0065.
\eeq
These correspond to the following observable parameters 
(at the \textit{pivot} scale) : 
\beq 
r = 0.101,~ n_s= 0.96, ~\frac{d n_s}{d \ln k} = 0.0111,~
~\frac{d^2 n_s}{d \ln k^2}  = 0.0139.
\label{eq:obs3a}
\eeq
\subitem (b)
In (A.$b$) we impose the criteria that at short length scales $(k > 10^{16} {\rm Mpc}^{-1})$ 
the scalar-power spectrum should be less than $10^{-2}$. This has also been imposed for all the 
other benchmarks. 
The slow-roll parameters at $\phicmb$ are given by, 
\beq
\varepsilon_{V*} = 0.0035, ~\eta_{V*}= -0.0095, 
~\xi_{V*}^2= -0.0073, ~\sigma_{V*}^3= 0.0021.
\eeq
These correspond to the following observable parameters 
(at the \textit{pivot} scale) : 
\beq 
r = 0.0056,~ n_s= 0.96, ~\frac{d n_s}{d \ln k} = 0.0138,~
~\frac{d^2 n_s}{d \ln k^2}  = 0.00486.
\label{eq:obs3b}
\eeq

\item {\bf Benchmark B} : (n = 4 case) \\
The parameters of the potential has been specified in row 3 of Table \ref{tab:bm}. 
The slow-roll parameters at $\phicmb$ are given by, 
\beq
\varepsilon_{V*} = 0.0045, ~\eta_{V*}= -0.0065, 
~\xi_{V*}^2= 0.0057, ~\sigma_{V*}^3= 0.0169.
\eeq
These correspond to the following observable parameters 
(at the \textit{pivot} scale) : 
\beq 
r = .072,~ n_s= 0.96, ~\frac{d n_s}{d \ln k} = -0.0124,~
~\frac{d^2 n_s}{d \ln k^2}  = 0.033.
\label{eq:obs4}
\eeq

\item {\bf Benchmark C} : (n = 6 case)\\  
The parameters in the potential has been mentioned in row 4 of Table \ref{tab:bm}.
At $\phi = \phicmb $, we have, 
\beq
\varepsilon_{V*} = 0.0028, ~\eta_{V*}= -0.0116, 
~\xi_{V*}^2= -0.0042, ~\sigma_{V*}^3= 0.018.
\eeq
These correspond to the following observable parameters 
(at the \textit{pivot} scale) : 
\beq 
r = 0.0448,~ n_s= 0.96, ~\frac{d n_s}{d \ln k} = 0.0077,~
~\frac{d^2 n_s}{d \ln k^2}  = 0.036.
\label{eq:obs6}
\eeq
\end{itemize}

For the above mentioned benchmarks points, we have $H \sim  10^{-5} M_P$; the variation of $H$ during inflation is typically 
by a factor of few. For $n=4$ and $n=6$, this corresponds to $\lambda_n \sim\mathcal{O}(10^{-6})$; 
and both $c_H$ and $a_H$ are of $\mathcal{O} (0.1)- \mathcal{O}(1)$. While for $n=3$ case, 
assuming $\lambda_3 \sim 10^{-6}$ would also yield similar values for $c_H$ and $a_H$. 
However, $\lambda_3$ is the coefficient of the superpotential term $N H_u.L$. Now, for such a large Yukawa
for $NH_uL$ we may not be able to assume the observed neutrino masses to be only Dirac type, which requires $\lambda_3\sim 10^{-12}$~\cite{Allahverdi:2006cx}. 
However, such a large Yukwas of $\lambda_3\sim 10^{-6}$ can yield the observed neutrino mass correct if we were to assume Majorana neutrino.
Needless to say that the above potential under consideration, see Eq.~(\ref{eq:pot1},~\ref{eq:pot}), can also accommodate Majorana type neutrinos but one 
with a soft mass term for the Majorana fields. If the Majorana masses for $\widetilde N$ are smaller than the Hubble parameter during inflation, then there would be 
no modification to our analysis, and the potential could be recast as in Eq.~(\ref{eq:pot1},~\ref{eq:pot}).

In all these cases the potential, which has been optimized to produce large $r$ ($r_*$) at the \textit{pivot} scale, 
shows a little bump around $\phicmb$. For $\phi > \phicmb$, the potential can rise quite fast 
for $n=6$ case, while it is more well-behaved for $n=3,4$. A sharp drop of the potential at
about $\phicmb$ leads to a large first derivative, which gives rise to a large $\varepsilon_{V*}$. 

The slow-roll parameters for $n=3$ case have been shown in Fig. (\ref{fig:1}). The left panel 
shows the evolution of $\varepsilon_V$ for both benchmarks (A.$a$) and (A.$b$). The right panel 
shows all slow-roll parameters only for (A.$a$). The same for  $n=4$ and $n=6$ cases have been 
plotted in Figs. (\ref{fig:1a}), (\ref{fig:1b}) respectively. The non-monotonic behavior of 
$\varepsilon_V$, and the peak of the same at $\phicmb$ further reflects on the shape of the 
potential. Further, soon after crossing $\phicmb$, $\varepsilon_V$ reaches its maximum, 
and then decreases rapidly assuring that slow-roll inflation continues. Clearly $\varepsilon_{V}$ 
around $\phicmb$ is most well-behaved for $n=3$ case, moderately so in $n=4$ case and shows rather 
sharp features for $n=6$ case. These features appear not only for our benchmarks, but also for 
most values of $\{V_0,~\varepsilon_{V*}\}$ producing large $r$ obtained in our scan.

Consequently, at the \textit{pivot} scale, the higher derivatives of 
the potential, and therefore the higher slow-roll parameters, remains most well-behaved for $n=3$ 
case. However, for $n=3$, the peak of $\varepsilon_{V}$ is rather broad, and it begins to descend 
fast for small $\phi$ compared to $n=4$ and $n=6$ case. This leads to a large $\eta_V$ at smaller 
$\phi$ values in this case. As we will see in the next section, large values of $\eta_V$ at small $\phi$ 
lead to more power at small length scales. Consequently, our requirement ${\cal P}_s (k \sim 10^{16} {\rm Mpc}^{-1}) 
\lesssim 10^{-2}$ constrains the maximum $r$ which can be obtained in this scenario. For example, 
benchmark (A.$a$) ( with $r \sim 0.1$) satisfies all CMB constraint at \textit{pivot} scale, but 
produces ${\cal P}_s \sim {\cal O}(1)$ at $k \sim 10^{16}  {\rm Mpc}^{-1}$. 
A similar situation applies to the $n=4$ case, while for $n=6$ the CMB constraints at the \textit{pivot}
plays a major role in constraining $r$ at the \textit{pivot} scale. 

Note that, although we require $N=50$, contributions to the energy density from the $\phi$ dependent 
part of the potential is significant only for about $10-15$ efoldings of inflation; following which 
$V_0$ dominates. After $50$ efoldings, therefore, some other mechanism is required to enhance the 
slow-roll parameters to terminate the inflation. This could be achieved via hybrid mechanism, or via
fast tunnelling to the true vacuum, as discussed in Refs.~\cite{Mazumdar:2011ih,Choudhury:2014sxa}.
As already mentioned, the non-monotonic behavior of $\varepsilon_V$ (and other slow-roll parameters) 
during the observational window of $8$ efoldings typically reflects on the scale dependence of 
the scalar and tensor power spectrum. In the next section we will address this issue in 
some detail.  

\section{Scale Dependence of the Perturbation}
\label{sec:pert}
\begin{figure}[t!]
\epsfig{file=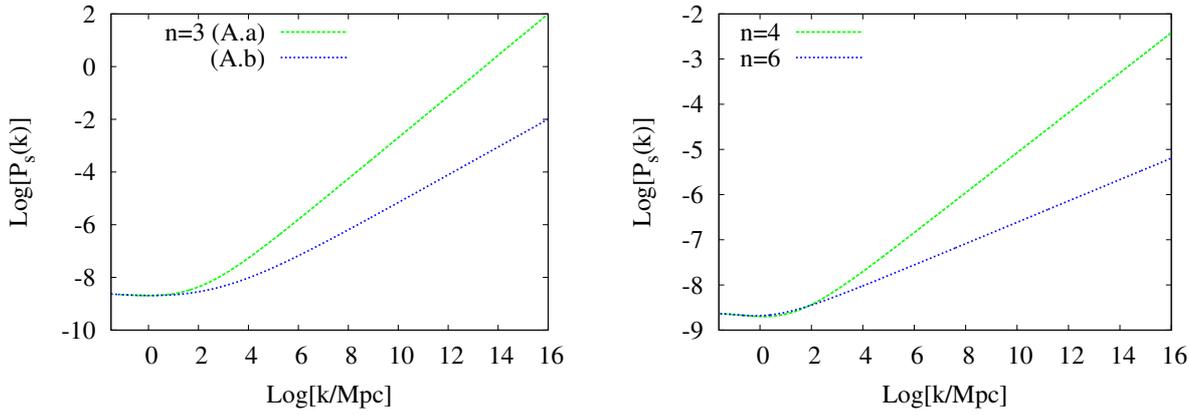, width= \textwidth}
\caption{The power-spectrum for scalar perturbations, ${\cal P}_s (k)$
has been plotted as a function of $\log \left(\frac{k}{\rm Mpc}\right)$
by evolving the mode equations numerically for the perturbations using the Sasaki-Mukhanov variable.
The left panel shows the variation of ${\cal P}_s$ with $k$ for benchmark scenarios (A$a$) and (A$b$) and the 
right panel shows the same for scenarios (B) and (C), see Table 2.}
\label{fig:3}
\end{figure}

\begin{figure}[ht!]
\epsfig{file=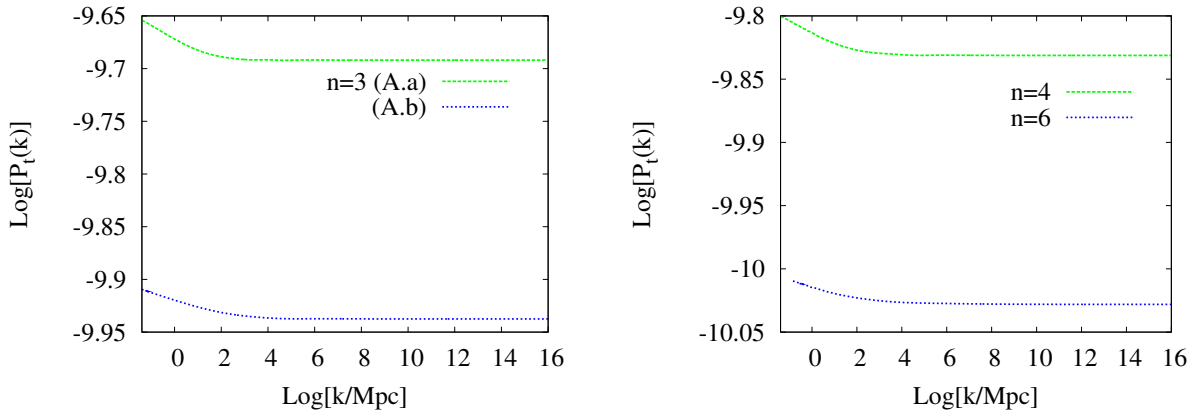, width= \textwidth}
\caption{The power-spectrum for tensor perturbations, ${\cal P}_t (k)$
has been plotted as a function of $\log \left(\frac{k}{\rm Mpc}\right)$ by evolving the mode 
equations numerically for the perturbations using the Sasaki-Mukhanov variable. The left panel 
shows the variation of ${\cal P}_t$ with $k$ for benchmark scenarios (A$a$) and (A$b$) and the 
right panel shows the same for scenarios (B) and (C), see Table 2.}
\label{fig:4}
\end{figure}

As shown in these figures, Figs.~(\ref{fig:3},\ref{fig:4}), both the scenarios are consistent with the CMB constraints at the 
\textit{pivot} scale and in the observational window. Also, except for scenario (A.$a$), 
all other benchmarks produce ${\cal P}_s(k=10^{16}{\rm Mpc}^{-1}) <10^{-2}$ at smaller length scales. In case of 
benchmark scenario (A.$a$) with $n=3$, large $\eta$ at small values of $\phi$, and 
correspondingly at small length scales, leads to an increase in the power. This is precisely the issue which 
was ignored in earlier discussions, especially in Refs.~\cite{Choudhury:2014kma}, and therefore it was possible 
to boost large $r$ at the pivot scale. Furthermore, the analytical treatment holds good for large scales close to the CMB 
scales, but breaks down as we extrapolate its evolution at the small scales.

The scale dependence of the tensor perturbation ${\cal P}_t (k)$ has been shown in Fig.~(\ref{fig:4}). Again, the left panel of Fig.~(\ref{fig:4}) demonstrates the scale 
dependence of ${\cal P}_t$ for benchmark scenarios (A.$a$) and (A.$b$), and the right panel shows 
the same for scenarios (B) and (C). The fall of the first slow-roll parameter about the 
pivot scale has been reflected in each of the sub figures. ${\cal P}_t(k)$ falls as $k$ increases 
past the pivot scale, and finally settles to a smaller value.  
Both of our benchmark scenarios respects  \texttt{PLANCK} constraints on ${\cal P}_s(k)$ at the 
pivot scale $k_{pivot}$, thanks to the potential reconstruction. However, non-monotonic 
evolution of the slow-roll parameters within the observational window is required to generate 
large $r_*$ evading the Lyth bound. As a consequence of the significant evolution of $\varepsilon_V$,  
during the Hubble crossing of the observable scales, significant scale dependent \textit{running} 
can be expected. This can result in deviation from simple power law dependence, 
and may produce large power at smaller length scales. As pointed out in  Ref.~\cite{HMN}, this may lead 
to a different value of $\sigma_8$ from the LSS data compared to that from CMBR data 
assuming a power law spectrum. Moreover, large power at smaller length scales ($k > k_{pivot}$) 
could lead to overproduction of small black holes, and may lead to a structure formation different 
from what has been observed. To avoid this, we require ${\cal P}_s(k=10^{16}{\rm Mpc}^{-1})< 10^{-2}$ at small length 
scales \cite{CKSY,DE1}. Therefore, we compute the power spectrum and reflect on its scale dependence 
by numerically solving the equations involved.  

To obtain ${\cal P}_s (k)$, we first solve the equation of motion for the inflation $vev$ ($\phi$), 
together with the Friedmann equations. These give the evolution of $\phi$ and the Hubble 
parameter $H$ during the inflationary epoch. After solving for the background evolution,   
we solve the Mukhanov-Sasaki equation \cite{MFB} numerically, without assuming the slow-roll 
condition, to obtain the mode functions corresponding to the scalar perturbation. We assume 
Bunch-Davies vacuum for these modes, and evolve these from sufficiently inside the Hubble to 
well after their Hubble crossing. The following Fig.~(\ref{fig:3}) demonstrates the scale 
dependence of ${\cal P}_s$ in the benchmark scnarios. The left panel of Fig.~(\ref{fig:3}) 
demonstrates the scale dependence of ${\cal P}_s$ for benchmark scenarios (A.$a$) and (A.$b$), 
while the right panel shows the same for scenarios (B) and (C). Similar plots were obtained by
evolving numerically the tensor perturbations, which has been shown in Figs.~(\ref{fig:4}).

Here we may speculate on the physics at short distances which might trigger formation of 
primordial black holes. Certainly, we know very little about how gravity is modified below 
$\sim 10^{-6}$m or energies higher than $\sim 10$~eV~\cite{Kapner:2006si}. The 
gravity itself could be modified beyond these energies in such a way that the gravitational 
interaction weakens as shown in Ref.~\cite{Biswas:2011ar}, effectively gravity becomes 
asymptotically free in such theories, and Schwarzschild's singularity can be avoided for a mass
 of black hole $m \lesssim M_P^2/M$. For $M\sim 10$~eV where gravity can be modified beyond this scale, $m\lesssim 10^{21}$g objects will
 never collapse to form black holes. Scenarios like this may completely avoid the PBH bounds on
 short scale physics. The question one may still ask - given our current treatment, is it necessary to 
 invoke such a scenario?, the outright answer would be no. Any modification of gravity would also lead to
 modifying the Sasaki-Mukhanov variable and would require reanalysis of the perturbation theory,
 which is beyond the scope of this current paper.

\section{Conclusion}
\label{sec:concl}
In this article we considered generation of large tensor-to-scalar ratio with sub-Planckian 
inflation model, focusing on the SUSY flat direction potential with large SUGRA corrections. 
We considered $3$ types of potentials, one renormalisable and $2$ non-renormalisable potentials
and scanned the model parameters which would satisfy the present constraints 
from \texttt{Planck}, and constraints arising on small scales primarily due to PBH 
abundance, which yields the ${\cal P}_s (k \sim 10^{16} {\rm Mpc}^{-1}) \lesssim 10^{-2}$.

The presence of large vacuum energy during inflation due to hidden sector physics and the
presence of {\it inflection point} allow non-monotonic evolution of the slow roll parameters.
In particular, the evolution of $\eta_V$ helps to boost the value of tensor-to-scalar ratio: $r\sim 0.1$.  
For the non-renormalisable operator $n=4$, see Eq.~(\ref{eq:pot1}), we obtained the largest value of 
$r\sim 0.072$, for $n=3$, $r=0.056$ and $n=6$, we obtained $r=0.045$. We have shown that due to non-monotonic 
evolution of slow roll parameter, although at small scales the power spectrum generally becomes large, 
it is consistent with the present data from LSS and CMB. 

This paper illustrates that if the future value of $r$ holds close to $r\sim 0.1$, it is quite possible to construct 
particle physics motivated models of inflation which are driven at sub-Planckian VEVs.  The inflaton candidates are well motivated 
$D$-flat directions of supersymmetric Standard Model,
such as $\widetilde N H_u\widetilde L,~H_uH_d,~\widetilde L\widetilde L\widetilde e,~\widetilde u\widetilde d\widetilde d$.
 Now the primary advantage 
is that one can account for all possible quantum corrections within an effective field theory assuming that the 
$4$ dimensional $M_P$ is the ultimate cut-off of the theory. Furthermore, since the inflaton candidates are made up of visible
sector fields, their decay naturally create the baryonic and the dark matter content of the observable universe.

\section{Acknowledgements}
AM is supported by the Lancaster-Manchester-Sheffield Consortium for Fundamental Physics under STFC grant ST/J000418/1.


\end{document}